\documentclass{aastex63}

\usepackage{amsmath}

\shortauthors{Gandolfi, Lapi, Liberati}
\shorttitle{Equilibria of non-minimally coupled DM halos}

\begin{document}

\title{Self-gravitating Equilibria of Non-minimally Coupled Dark Matter Halos}

\author[0000-0003-3248-5666]{Giovanni Gandolfi}\affiliation{SISSA, Via Bonomea 265, 34136 Trieste, Italy}\affiliation{IFPU - Institute for fundamental physics of the Universe, Via Beirut 2, 34014 Trieste, Italy}

\author[0000-0002-4882-1735]{Andrea Lapi}
\affiliation{SISSA, Via Bonomea 265, 34136 Trieste, Italy}\affiliation{IFPU - Institute for fundamental physics of the Universe, Via Beirut 2, 34014 Trieste, Italy}\affiliation{INFN-Sezione di Trieste, via Valerio 2, 34127 Trieste,  Italy}\affiliation{INAF-Osservatorio Astronomico di Trieste, via Tiepolo 11, 34131 Trieste, Italy}

\author[0000-0002-7632-7443]{Stefano Liberati}\affiliation{SISSA, Via Bonomea 265, 34136 Trieste, Italy}\affiliation{IFPU - Institute for fundamental physics of the Universe, Via Beirut 2, 34014 Trieste, Italy}\affiliation{INFN-Sezione di Trieste, via Valerio 2, 34127 Trieste,  Italy}

\begin{abstract}
We investigate self-gravitating equilibria of halos constituted by dark matter (DM) non-minimally coupled to gravity.  In particular, we consider a theoretically motivated non-minimal coupling which may arise when the averaging/coherence length $L$ associated to the fluid description of the DM collective behavior is comparable to the local curvature scale. In the Newtonian limit, such a non-minimal coupling amounts to a modification of the Poisson equation by a term $L^2\,\nabla^2\rho$ proportional to the Laplacian of the DM density $\rho$ itself. We further adopt a general power-law equation of state $p\propto \rho^{\Gamma}\, r^\alpha$ relating the DM dynamical pressure $p$ to density $\rho$ and radius $r$, as expected by phase-space density stratification during the gravitational assembly of halos in a cosmological context. We confirm previous findings that, in absence of the non-minimal coupling, the resulting density $\rho(r)$ features a steep central cusp and an overall shape mirroring the outcomes of $N-$body simulations in the standard $\Lambda$CDM cosmology, as described by the classic NFW or Einasto profiles. Most importantly, we find that the non-minimal coupling causes the density distribution to develop an inner core and a shape closely following, out to several core scale radii, the Burkert profile. In fact, we highlight that the resulting mass distributions can  fit,  with  an  accuracy  comparable  to the  Burkert's one, the co-added  rotation curves of dwarf, DM-dominated galaxies. Finally, we show that non-minimally coupled DM halos are consistent with the observed scaling relation between the core radius $r_0$ and core density $\rho_0$, in terms of an universal core surface density $\rho_0\times r_0$ among different galaxies.
\end{abstract}

\keywords{Cosmology (343) - Dark matter (353) - Non-standard theories of gravity (1118)}

\section{Introduction}\label{sec|intro}

$N-$body, dark matter (DM)-only simulations in the standard $\Lambda$CDM cosmology suggest an almost universal shape of the density distributions $\rho(r)$ within virialized DM halos of different masses and redshifts. This fact has been established since the seminal work by Navarro et al. (1996), who proposed an empirical fitting formula (the so-called NFW profile) for the DM density run $\rho(r)\propto \frac{1}{r\,(r+r_s)^2}$ that still nowadays constitute the standard lore (but see Navarro et al. 2010 for refinements); inward of the scale radius $r_s$ a density cusp, i.e. a steep central divergence, is found.

However, such a behavior is at variance with the observational evidences inferred from well-measured rotation curves in many dwarf, DM-dominated galaxies, that point toward a constant, finite inner density $\rho_0$ within a core radius $r_0$ (see, e.g., McGaugh et al. 2001; Gentile et al. 2004; de Blok et al. 2008; Walker \& Penarrubia 2011; Weinberg et al. 2015; Genzel et al. 2020; for a review and further references, see Bullock \& Boylan-Kolchin 2017). The observed density distribution is usually described with the phenomenological Burkert (1995) profile $\rho(r)\propto \frac{1}{(r+r_0)\,(r^2+r_0^2)}$.

A non-trivial point to stress is that, besides the flat shape of the inner density run, the measurements indicate a nearly universal relationship between the core density and radius, in terms of a constant value for the ``core surface density" $\rho_0\times r_0$ among different galaxies (see Salucci \& Burkert 2000; Donato et al. 2009; Gentile et al. 2009; Burkert 2015; Kormendy \& Freeman 2016). In addition, another puzzling aspect of the DM phenomenology is the existence of tight scaling laws with baryonic quantities. These include the baryonic Tully-Fisher relation (Tully \& Fisher 1977; McGaugh 2012), the core radius vs. disc scale-length relation (see Donato et al. 2004), the radial acceleration relation (see McGaugh et al. 2016), and the universal DM-baryon constant (see Chan 2019).

Admittedly, through the last $20$ years the cusp-core controversy has flamed the scientific debate. One class of solutions invokes physical processes that can cause violent fluctuations in the inner gravitational potential and/or transfer of energy and angular momentum from the baryons to DM, thus possibly erasing the central density cusp; some of the most explored possibilities include dynamical friction (see El-Zant et al. 2001, 2016; Tonini et al. 2006; Romano-Diaz et al. 2008), and feedback effects from stars and active galactic nuclei (see Governato et al. 2012; Teyssier et al. 2013; Pontzen \& Governato 2014; Peirani et al. 2017; Freundlich et al. 2020a). In particular, hydrodynamic simulations including baryonic physics have shown that a variety of halo responses are originated depending on the stellar/halo masses (e.g., Freundlich et al. 2020b).

An alternative solution, perhaps more fascinating, is to abandon the cold DM hypothesis and look at non-standard particle candidates (see review by Salucci 2019). A few examples include: self-interacting DM particles with cross-section $\sim 1$ cm$^2$ g$^{-1}$, which can create a core as they are heated up via elastic two-body collisions and evacuated from the inner region  (e.g., Spergel \& Steinhardt 2000; Vogelsberger et al. 2014); a self-interacting or non-interacting (alias fuzzy DM) Bose-Einstein condensate of ultra-light particles (likely axions) with masses $\sim 10^{-22}$ eV, for which the core stems from the equilibrium between quantum pressure and gravity (e.g., Hu et al. 2000; Bohmer \& Harko 2007; Schive et al. 2014a,b; Harko 2014; Hui et al. 2017; Bernal et al. 2018); warm DM, made of fermionic particles with masses of order a few keV (e.g., sterile neutrinos), that can originate cores where gravity is counterbalanced by quantum degeneracy pressure from the Pauli exclusion principle (e.g., Dodelson \& Widrow 1994; Shi \& Fuller 1999; Kusenko 2009; Destri et al. 2013; Adhikari et al. 2017). A more radical perspective circumventing the core-cusp problem envisages that no DM is present and tries to explain the galactic dynamics via a modification of gravity, like in the MOND phenomenological approach (see Milgrom 1983, 2009; Bruneton \& Esposito-Farese 2007; Bekenstein 2004, 2009; for a review and further references, see Famaey \& McGaugh 2012).

Here we take yet another viewpoint, retaining standard cold DM but envisaging that its dynamics may be subject to a non-minimal coupling with gravity (see Bruneton et al. 2009; Bertolami \& Paramos 2010; Bettoni et al. 2011; Bettoni \& Liberati 2015; Ivanov \& Liberati 2020). The  rationale of such attempts is trying to keep the collisionless DM phenomenology on large cosmological scales, and at the same time to introduce MOND-like behavior in galaxies by attributing to DM some non-negligible coupling with gravity. The words ``non-minimal" simply mean that such a coupling is embodied in the action via an additional interaction term with respect to standard scalar-tensor gravity theories. We stress that the non-minimal coupling is not necessarily a fundamental feature of the DM particles but might develop dynamically when the averaging/coherence length $L$ associated to the the fluid description of the matter collective behavior is comparable to the local curvature scale. We specifically consider and theoretically justify a form of such a coupling that in the Newtonian limit amounts to a modification of the Poisson equation by a term $L^2\,\nabla^2\rho$ proportional to the Laplacian of the DM density $\rho$ itself, thus expressing an effective coupling of the DM fluid with the local gravitational curvature. Note that extensions of gravity including non-minimal couplings have been also investigated on cosmological scales, since under certain conditions they can mimic and explain the properties of the dark energy component (e.g., Bettoni et al. 2012; Bertolami \& Paramos 2014).

In the present work, we show that non-minimally coupled DM halos in self-gravitating equilibria feature a density distribution closely following the Burkert profile out to several core scale radii. Moreover, we show that these can fit very well
the measured rotation curves of dwarf, DM-dominated galaxies, and are consistent with the observed universal core surface density. In more detail, the plan of the paper is the following: in Sect.~\ref{sec|NMC-DM} we theoretically motivate the adopted non-minimally coupled DM framework; in Sect.~\ref{sec|basics} we introduce the basic formalism to describe self-gravitating equilibria of non-minimally coupled DM halos; in Sect.~\ref{sec|EOS} we describe the effective equation of state for the DM fluid; in Sect.~\ref{sec|fundeq} we derive the fundamental equation ruling the DM density profile and study its solution space; in Sect.~\ref{sec|profcomp} we compare our non-minimally coupled solutions with the density profiles classically adopted in the literature to fit simulations and/or observations; in Sect.~\ref{sec|datacomp} we provide a first glimpse on how our non-minimally coupled DM mass distributions can fit the measured rotation curves of dwarf, DM-dominated galaxies; in Sect.~\ref{sec|scaling} we show that non-minimally coupled DM halos are indeed consistent with the observed universal behavior of the core surface density; finally, in Sect.~\ref{sec|summary} we summarize our findings and highlight future prospects.

Throughout this work, we adopt the standard flat $\Lambda$CDM cosmology (Planck Collaboration 2020) with rounded parameter values: matter density $\Omega_M=0.3$, dark energy density $\Omega_\Lambda=0.7$, baryon density $\Omega_b=0.05$, and Hubble constant $H_0 = 100\, h$ km s$^{-1}$ Mpc$^{-1}$ with $h=0.7$; unless otherwise specified, $G\approx 6.67\times 10^{-8}$ cm$^3$ g$^{-1}$ s$^{-2}$ indicates the standard gravitational (Newton) constant.

\section{A theoretical framework for non-minimally coupled DM}\label{sec|NMC-DM}

The motivation for introducing a non-minimal coupling between the DM matter field and gravity is twofold. On the theoretical side, it is allowed by the Einstein equivalence principle (Di Casola et al. 2015), and it might be required for the renormalizability of quantum field theories in curved spacetimes (e.g., Sonego \& Faraoni 1993; Bruneton et al. 2009). On the observational side, it may help in explaining the striking relationships between DM and baryons recalled in Sect.~\ref{sec|intro}
(e.g., baryonic Tully-Fisher, core radius vs. disc scale-length, acceleration relations, etc.) that are not trivially understood via galaxy formation processes. In fact, such relationships may indicate a very special DM-baryon interaction. However, on the one hand, evidence of any direct coupling is missing. On the other hand, modified-gravity models like MOND seem, at least in galactic environments, able to describe the baryon dynamics as that of freely falling particle on some modified gravitational background.

Following these hints one can be led to conjecture, as in Bruneton et al. (2009), that the physical metric experienced by baryons may not coincide with the gravitational one, but it may be also determined by the properties of the DM field. Indeed, it was shown by Bekenstein (1993) that the most generic transformation between physical and gravitational metric preserving causality and the weak equivalence principle can be of the general form
\begin{equation}\label{eq|disformal}
\widetilde{g}_{\mu \nu} = e^{2\varphi} \left [\mathcal{A}(\mathcal{X}) g_{\mu \nu} + \mathcal{B}(\mathcal{X})\nabla_{\mu}\varphi \nabla_{\nu}\varphi \right]~,
\end{equation}
where $\mathcal{A}$ and $\mathcal{B}$ are functions to be specified, $\varphi$ is an extra scalar field and $\mathcal{X}=-\frac{1}{2}\,g_{\mu\nu}\nabla^\mu\varphi\nabla \partial^\nu \varphi$; such a relation between metrics is called a disformal one. Following the aforementioned idea, we can now ask which kind of interaction might be reduced to an effective coupling via a disformal metric of the above form with the scalar field playing the role of DM.

To this purpose, one can start from a general action of the form
\begin{equation}
S = S_{\rm EH}[g_{\mu \nu}] + S_{\rm bar}[g_{\mu \nu},\psi]+S_{\rm DM}[g_{\mu \nu},\varphi]+S_{\rm int}[g_{\mu \nu}, \psi,\varphi]~,
\end{equation}
where the terms on the right hand side identify respectively the Einstein-Hilbert (standard general relativity) action $S_{\rm EH}=\frac{c^4}{16\pi G}\,\int {\rm d}^{4}x\,\sqrt{-g}\, R$ in terms of the Ricci scalar $R$, the baryonic $S_{\rm bar}$ and DM $S_{\rm DM}$ actions, and the interaction one $S_{\rm int}$. Here the scalar fields $\psi$ and $\varphi$ are thought as collective variables encoding baryons and DM, respectively. It is then easy to show that DM can produce an effective metric for the baryons of the form specified by Eq.~(\ref{eq|disformal}) if $S_{\rm bar}[ g_{\mu \nu},\psi] + S_{\rm int}[g_{\mu \nu}, \psi,\varphi] \approx S_{\rm bar}[g_{\mu \nu}+h_{\mu \nu},\psi]$ with  $h_{\mu \nu}\propto \nabla_{\mu}\varphi \nabla_{\nu}\varphi$. Up to order $\mathcal{O}(h^{2})$, this is originated by an interaction term of the form $S_{\rm int} [g_{\mu \nu}, \psi,\varphi] \propto \int{\rm d}^{4}x\,\sqrt{-g}\,T^{\mu \nu}_{\rm bar}\,\nabla_{\mu}\varphi \nabla_{\nu}\varphi$, where $T^{\mu \nu}_{\rm bar}$ is the baryonic matter stress-energy tensor.

Noticeably, if now we express the full action in terms of the physical metric $\widetilde{g}_{\mu \nu} \equiv g_{\mu \nu} + h_{\mu \nu}$ (which is equivalent to choose a frame in which baryons follow the geodesics of this metric, i.e., the Jordan frame), then it turns out that the DM field gets non-minimally coupled to gravity as
\begin{equation}\label{eq|jordan}
S = S_{\rm EH}[\widetilde{g}_{\mu \nu}] + S_{\rm bar}[\widetilde{g}_{\mu \nu},\psi]+S_{\rm DM}[\widetilde{g}_{\mu \nu},\varphi]
+\epsilon L^2\, \int {\rm d}^{4}x\,\sqrt{-\tilde g}\, \widetilde{G}^{\mu \nu}\, \nabla_{\mu}\varphi \nabla_{\nu}\varphi~,
\end{equation}
where $L$ is a coupling length that must be present for dimensional consistency, $\epsilon=\pm 1$ is a constant that represents the polarity of the coupling (undetermined a priori), $\widetilde{G}^{\mu \nu}$ is the Einstein tensor expressed in terms of the physical metric $\widetilde{g}_{\mu\nu}$, and the DM field has been implicitly redefined through a conformal factor.

Three remarks are in order. First, note that the only other non-minimal coupling term with the same physical dimensions (still leading to second order field equations) would be proportional to the Ricci scalar as $\mathcal{X}\,\widetilde{R}$, which is however equivalent to that appearing in Eq.~(\ref{eq|jordan}) modulo a surface term (see Bettoni \&  Liberati 2013). Second, the coupling term $\widetilde{G}^{\mu \nu}\, \nabla_{\mu}\varphi \nabla_{\nu}\varphi$ appearing in the above action is proportional to a term of the Horndeski Lagrangian, which constitutes the most general scalar tensor theory giving rise to second order field equations; however, here we are not proposing a fundamental theory of modified gravity, but just
entailing the possibility that DM in galactic halos dynamically develops a non-minimal coupling with the metric characterized by an effective length-scale $L$. Third, it can be shown that by choosing a special form of the tensor $h_{\mu\nu}$, one can reproduce the MONDian regime in galaxies (see  Bruneton et al. 2009, their Eqs. 2.6-2.7). In the present paper, we rely on the much simpler form $h_{\mu \nu}\propto \nabla_{\mu}\varphi \nabla_{\nu}\varphi$ as above, which will not provide a MONDian limit, but will produce kinematics consistent with observations (at least in dwarf, DM-dominated galaxies); from this point of view, the coupling length $L$ cannot be straightforwardly interpreted in terms of the acceleration parameter $a_0$ appearing in the MONDian dynamics.

In order to take the Newtonian limit of the above theory it is convenient to adopt fluid variables. In the case of a complex scalar field\footnote{While for simplicity we have written our formulas in terms of a real scalar field, the aforementioned derivative non-minimal coupling can be easily generalized to a complex scalar field as $\widetilde{G}^{\mu \nu}\nabla_{\mu}\varphi \nabla_{\nu}\varphi^\dagger$.} this is easily achieved by adopting the standard Madelung representation $\varphi\sim\sqrt{\rho} e^{i\theta}$. However, a fluid limit is possible also for a non-minimally coupled real scalar field (see Bettoni et al. 2012). In the end it can be shown that the non-minimal coupling considered above leads to a modified Poisson equation of the form (see e.g., Bettoni et al. 2014)
\begin{equation}\label{eq|modpoisson}
\mathbf{\nabla}^2\Phi =4\pi G\,[(\rho+\rho_{\rm bar}) - \epsilon\,L^2\,\nabla^2\rho]~,
\end{equation}
where $\Phi$ is the Newtonian potential, and $\rho_{\rm bar}$ and $\rho$ are the baryon and DM mass densities. This modified Poisson equation implies that the source for gravity is not just the total matter density in itself, but also the DM inhomogeneities or spatial variations\footnote{Remarkably, the same formal modification can be obtained starting directly from a cosmological fluid description (see Bettoni \& Liberati 2015) albeit in this case there is no fundamental reason to couple the fluid directly to the Einstein tensor (and indeed it is necessary to couple separately the fluid to the Ricci scalar and/or Ricci tensor to get the same kind of modification, see Bettoni et al. 2014). It is also worth reporting that the same form of modified Poisson equation can be derived in Born-Infield gravity (e.g., Beltran Jimenez et al. 2018).}. In the present paper we will focus mainly on dwarf, DM-dominated galaxies and thus we will neglect the baryonic component $\rho_{\rm bar}$ hereafter.

While the discussed non-minimal coupling could be associated to some modified theory of gravity, in the extant literature at least two main mechanisms have been contemplated to produce dynamically a non-minimal coupling characterised by a length-scale $L$: either it can emerge from some collective behavior of the DM particles associated to a coherence length (for example via a Bose-Einstein condensation mechanism; see Bettoni et al. 2011 for an extended discussion), or it appears through an averaging procedure associated to the fluid description of matter. Although in general we adhere to the latter viewpoint, discussing the physical emergence of $L$ is beyond the scope of this paper; indeed in what follows we shall investigate the implications of the modified Poisson Eq.~(\ref{eq|modpoisson}) for the self-gravitating equilibria of DM halos, without an {\em a priori} prejudice about the origin and scale of $L$.

\section{Self-gravitating equilibria of DM halos}\label{sec|basics}

The self-gravitating equilibria of DM halos can be specified in the fluid approximation (Teyssier et al. 1997; Subramanian et al. 2000; Lapi \& Cavaliere 2011; Nadler et al. 2017) via the continuity, Euler (also called Jeans in this context), and Poisson equations
\begin{equation}\label{eq|basics}
\left\{
\begin{aligned}
&\partial_t\,\rho+\mathbf{\nabla}\cdot (\rho\,\mathbf{v})=0~,\\
\\
&\partial_t\,\mathbf{v}+(\mathbf{v}\cdot \mathbf{\nabla})\,\mathbf{v}+\cfrac{1}{\rho}\, \mathbf{\nabla}\,p=-\mathbf{\nabla}\,\Phi~,\\
\\
&\mathbf{\nabla}^2\Phi =4\pi G\,(\rho - \epsilon\,L^2\,\nabla^2\rho)~.
\end{aligned}
\right.
\end{equation}
Here  $\mathbf{v}$ is the bulk velocity, $p=\rho\, \sigma_r^2$ is the pressure dynamically generated by the random motions (and specified in terms of a radial velocity dispersion $\sigma_r^2$ or more generally of an anisotropic stress tensor $\sigma_{ij}^2$) of the DM particles in approximate virial equilibrium within the gravitational potential well\footnote{Note that $p$ is not to be confused with the relativistic pressure $p\approx 0$ adopted for cold DM in a cosmological context.}. The last term on the right hand side of the Poisson equation represents the non-minimal coupling with length-scale $L$ and polarity $\epsilon$ as discussed in Sect.~\ref{sec|NMC-DM}; we will see that for the purpose of originating physically acceptable DM density distributions, a negative polarity $\epsilon=-1$ is required, while $L$ will turn out to be closely related to the DM core radius. Note that non-minimal coupling terms do not appear in the Euler equation since they are found to be sub-leading in the non-relativistic limit (i.e., expansion in $1/c^2$; see Bettoni et al. 2014).

\subsection{Equation of state}\label{sec|EOS}

To close the system, the pressure must be related to the density via an equation of state (EOS). In the present context, we focus on the EOS originated by the gravitational assembly of DM halos via accretion and mergers from the cosmic web. This EOS stems from the progressive stratification of the DM pseudo-entropy $K\equiv p/\rho^{5/3}$, or equivalently of the coarse-grained phase-space density $\rho/\sigma_r^3\propto K^{-3/2}$, in terms of a simple power-law profile $K(r)\propto r^\alpha$. Although the physical origin of this scale-free behavior is not fully understood (see Nadler et al. 2017; Arora \& Williams 2020), $N-$body simulations (Peirani et al. 2006; Navarro et al. 2010; Ludlow et al. 2011; Gao et al. 2012; Nolting et al. 2016; Butsky et al. 2016) have shown this to approximately hold over more than three order of magnitude in radius within virialized halos, with powerlaw index  $\alpha\approx 1.25-1.3$. Such values are indeed expected on the basis of simple self-similar arguments (see Bertschinger 1985; Lapi \& Cavaliere al. 2009a, 2011; Nadler et al. 2017) and also broadly consistent with observations (see Lapi \& Cavaliere 2009b; Chae 2014; Munari et al. 2014). On this basis, but to keep some degree of generality\footnote{For example, some authors (e.g., Schmidt et al. 2008; Hansen et al. 2010) have claimed that it is the quantity $\rho/\sigma_r^\varepsilon\propto r^{-\zeta}$ with $\varepsilon\lesssim 3$ to feature a powerlaw behavior; in our parametrization of Eq.~(\ref{eq|EOS}) this just amounts to take $\alpha=\frac{2}{3}\,\zeta$ and $\Gamma=1+\frac{2}{\varepsilon}$.}, we adopt the EOS parametrization
\begin{equation}\label{eq|EOS}
p(\rho,r) = \lambda\, \rho^{\Gamma}\, r^{\alpha}= \rho_\alpha\,\sigma^2_\alpha\,\left(\cfrac{\rho}{\rho_\alpha}\right)^\Gamma\,
\left(\cfrac{r}{r_\alpha}\right)^\alpha~,
\end{equation}
where $r_\alpha$ is a reference radius and we have defined $\rho_\alpha\equiv \rho(r_\alpha)$ and $\sigma^2_\alpha\equiv \sigma_r^2(r_\alpha)$; in the following we adopt as fiducial values $\Gamma\approx \frac{5}{3}$ and $\alpha\approx 1.3$.

Under static ($\mathbf{v}=0$), spherically symmetric, and isotropic conditions (see Appendix for a generalization) the relevant Eqs.~(\ref{eq|basics}) become
\begin{equation}\label{eq|static}
\left\{
\begin{aligned}
& \cfrac{1}{\rho}\,\cfrac{{\rm d}p}{{\rm d}r} = - \cfrac{{\rm d}\Phi}{{\rm d}r}~,\\
\\
& \cfrac{1}{r^2}\,\cfrac{{\rm d}}{{\rm d}r}\,\left(r^2\,\cfrac{{\rm d\Phi}}{{\rm d}r}\right) = 4\pi G\, \left[\rho-\epsilon\, L^2\, \cfrac{1}{r^2}\,\cfrac{{\rm d}}{{\rm d}r}\,\left(r^2\,\cfrac{{\rm d\rho}}{{\rm d}r}\right)\right]~,
\end{aligned}
\right.
\end{equation}
supplemented with the trivial mass conservation constraint $\mathcal{M}(t)=$const, where
\begin{equation}\label{eq|massconst}
\mathcal{M}\equiv 4\pi\int_0^\infty{\rm d}r\,r^2\,\rho(r)~,
\end{equation}
is the total mass $\mathcal{M}$ of the DM halo.

\begin{figure}
\centering
\includegraphics[width=0.8\textwidth]{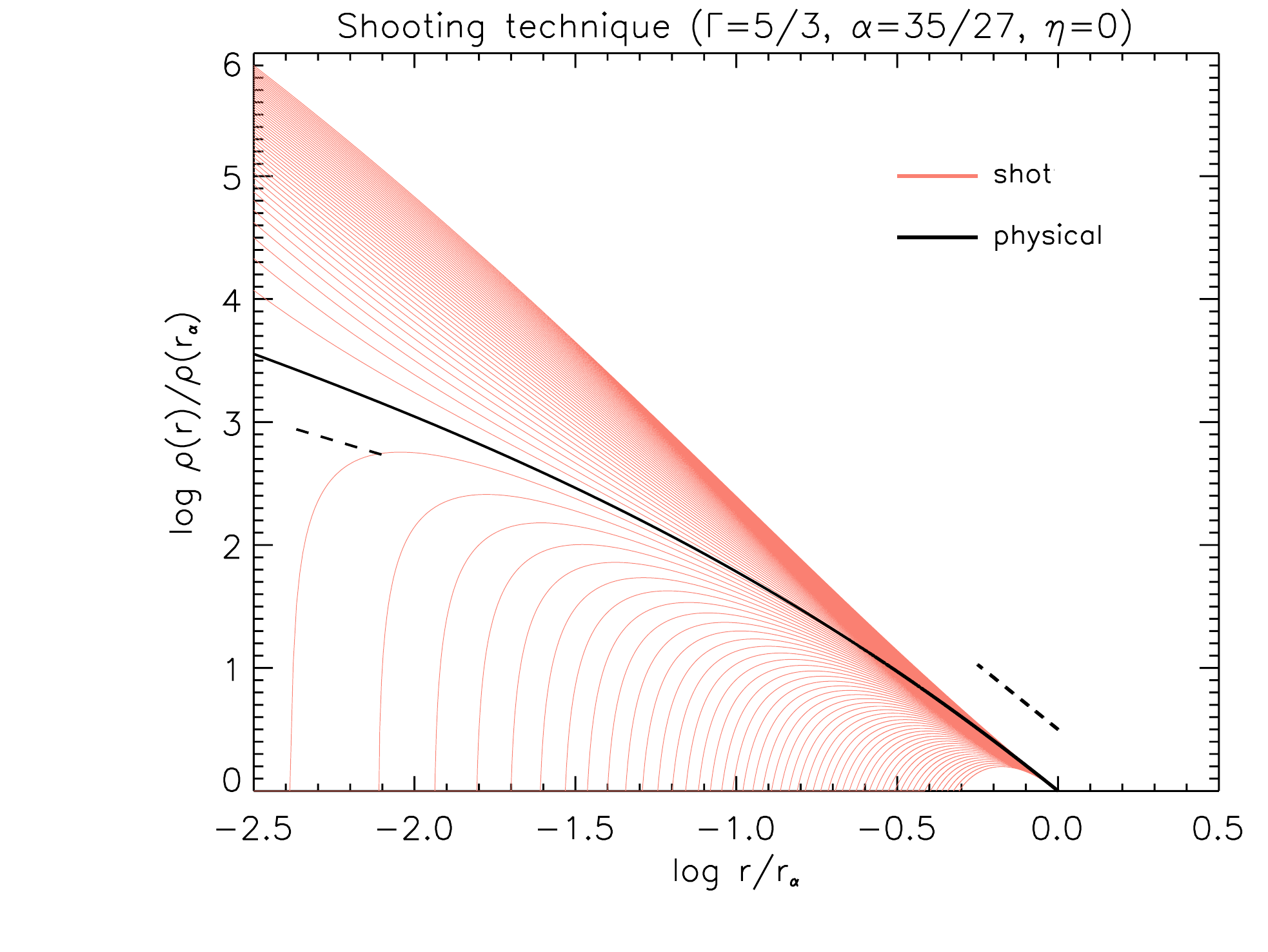}
\caption{Example aimed at illustrating the shooting technique adopted to solve the fundamental Eq.~(\ref{eq|fundeq}) and to find the physically acceptable solutions. Radial coordinate and density profile are normalized to the reference radius $r_\alpha$ where the logarithmic slope $\gamma\equiv-\frac{{\rm d}\log \rho}{{\rm d}\log r}$ of the profile is $\gamma_\alpha= \frac{2-\alpha}{2-\Gamma}$. The differential equation is integrated inward of $r_\alpha$ with boundary condition $\rho(r_\alpha)=\rho_\alpha$ and $\rho'(r_\alpha)=-\gamma_\alpha\, \frac{\rho_\alpha}{r_\alpha}$. Red lines show shot solutions for different values of the constant $\kappa$ appearing in Eq.~(\ref{eq|fundeq}), while the black line is the only physical profile for $\kappa=\kappa_{\Gamma,\alpha}$, see text for details (for $\kappa>\kappa_{\Gamma,\alpha}$ shot solutions are below the physical profile, while for $\kappa<\kappa_{\Gamma,\alpha}$ are above). The two  dashed lines indicate the asymptotic slope $\gamma_0=\frac{\alpha}{\Gamma}$ in the inner region and the slope $\gamma_\alpha$ at $r=r_\alpha$. In this example we specifically considered the minimally-coupled case $\eta=0$, and EOS parameters $\Gamma=\frac{5}{3}$ and $\alpha=\alpha_{\rm crit}=\frac{35}{27}$, yielding $\kappa_{\Gamma,\alpha}\approx 2.5$,  $\gamma_0=\frac{7}{9}$ and $\gamma_\alpha=\frac{19}{9}$.}\label{fig|NMC_shooting}
\end{figure}

\begin{figure}
\centering
\includegraphics[width=0.8\textwidth]{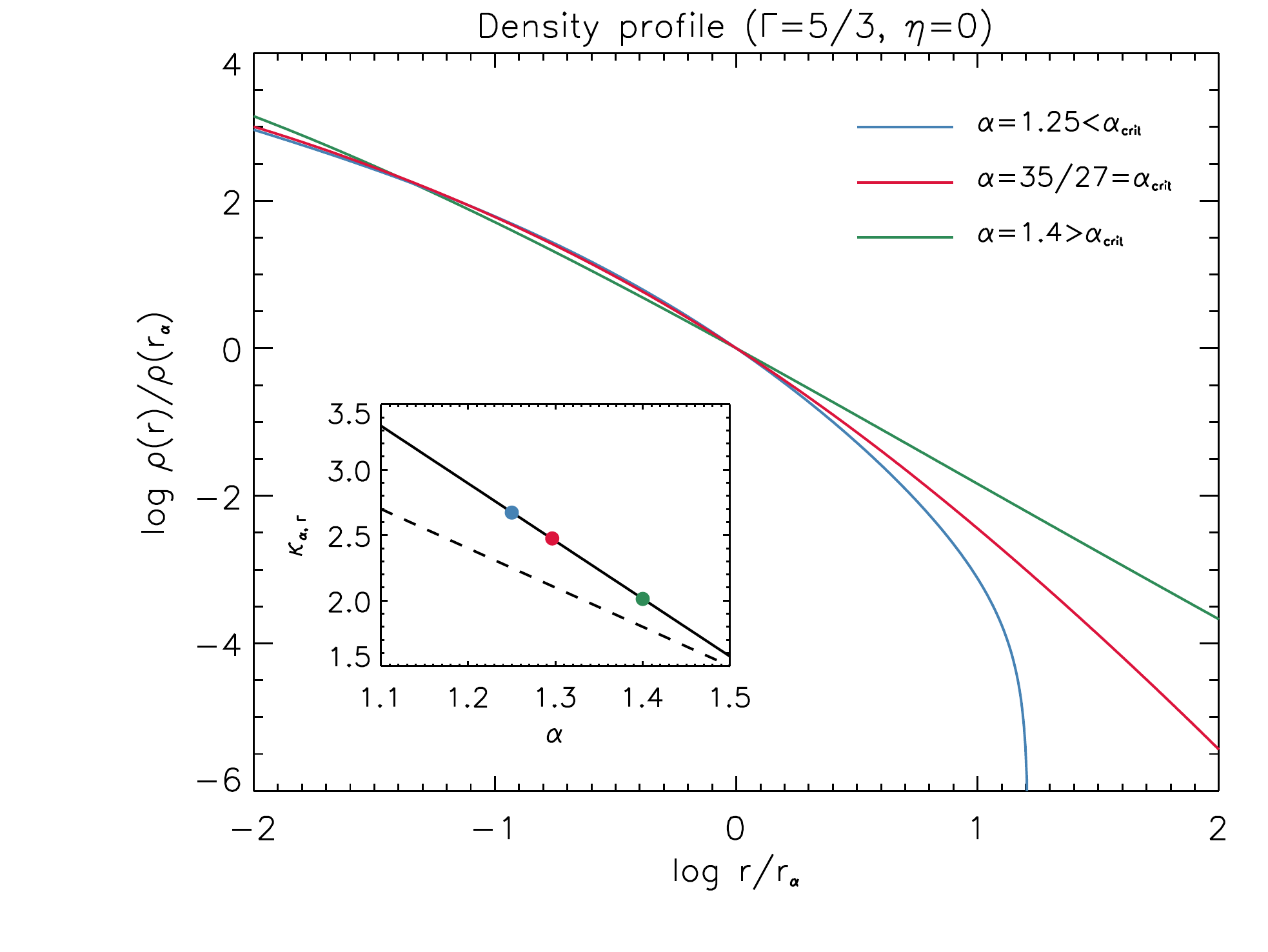}
\caption{Examples aimed at showing the outer behavior of the density profiles from solving the fundamental Eq.~(\ref{eq|fundeq}). The illustrated profiles refer to the minimally coupled case $\eta=0$ with EOS parameter $\Gamma=\frac{5}{3}$ and three different values of $\alpha$: the red line corresponds to $\alpha=\frac{35}{27}=\alpha_{\rm crit}$ for which the density asymptotes to a slope $\gamma_\infty=\frac{\alpha+1}{\Gamma-1}\approx \frac{31}{9}$; the blue line refers to $\alpha=1.25<\alpha_{\rm crit}$ for which the slope $\gamma_\infty$ is attained at a finite radius before a cutoff; the green line refers to $\alpha=1.4>\alpha_{\rm crit}$ for which the outer slope is unphysical since it would lead to a diverging mass. The inset reports the dependence on $\alpha$ of the constant $\kappa_{\Gamma,\alpha}$ for the full solutions (colored dots and solid black line), and for the pure power-law solutions (dashed black line), see Sect.~\ref{sec|fundeq} for details.}\label{fig|NMC_density_alpha}
\end{figure}

\begin{figure}
\centering
\includegraphics[width=0.8\textwidth]{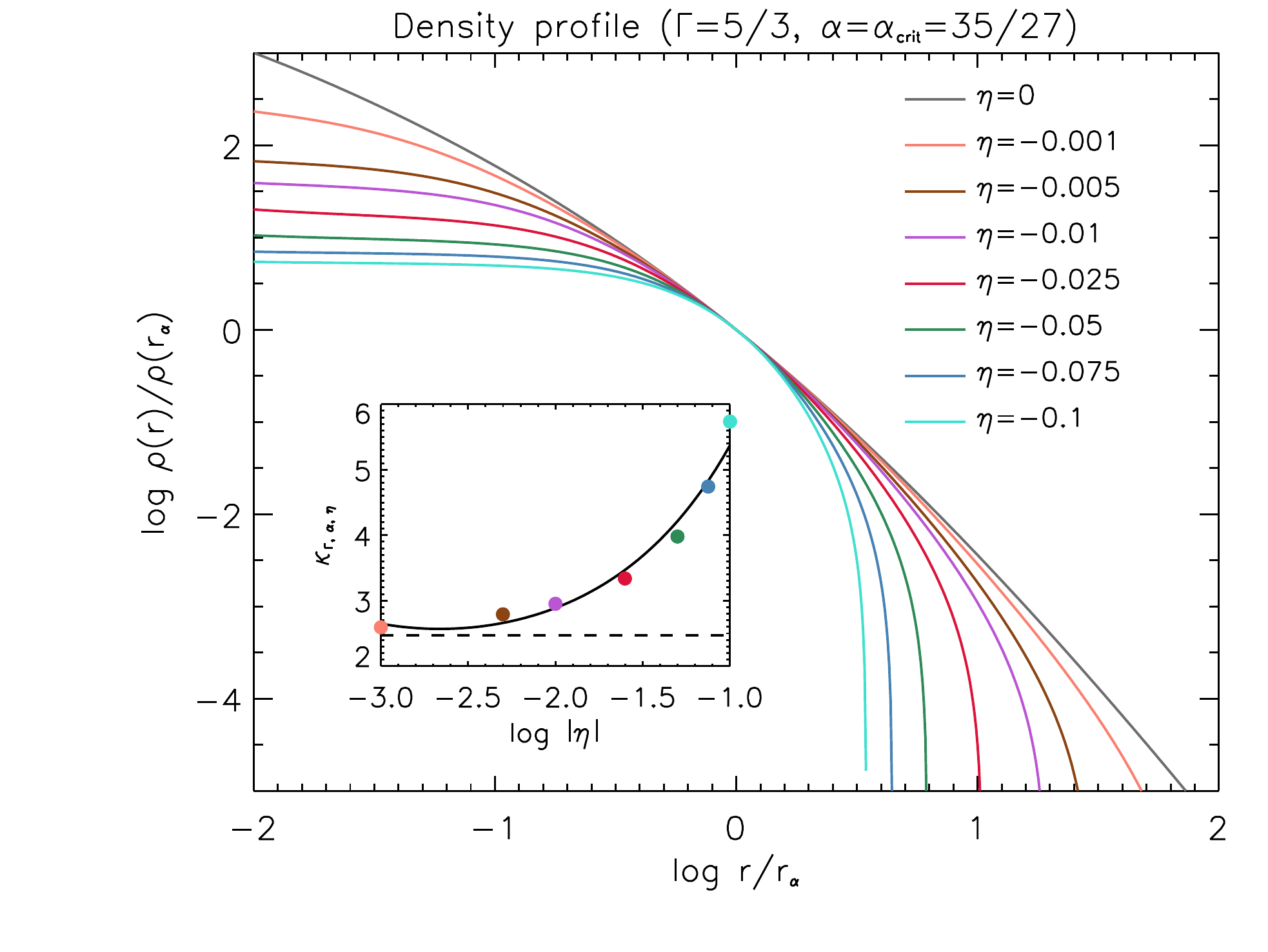}
\caption{Density profiles (EOS parameters $\Gamma=\frac{5}{3}$ and $\alpha=\alpha_{\rm crit}=\frac{35}{27}$) for different values of the non-minimal coupling $\eta=0$ (grey), $-0.001$ (orange), $-0.005$ (brown), $-0.01$ (magenta), $-0.025$ (red), $-0.05$ (green), $-0.075$ (blue), $-0.1$ (cyan). The inset reports the corresponding values of the constant $\kappa_{\Gamma,\alpha,\eta}$ for the physical solutions as a function of $\eta$ (colored dots and solid line), and for the $\eta=0$ case (dashed black line).}\label{fig|NMC_density_eta}
\end{figure}

\begin{figure}
\centering
\includegraphics[width=0.8\textwidth]{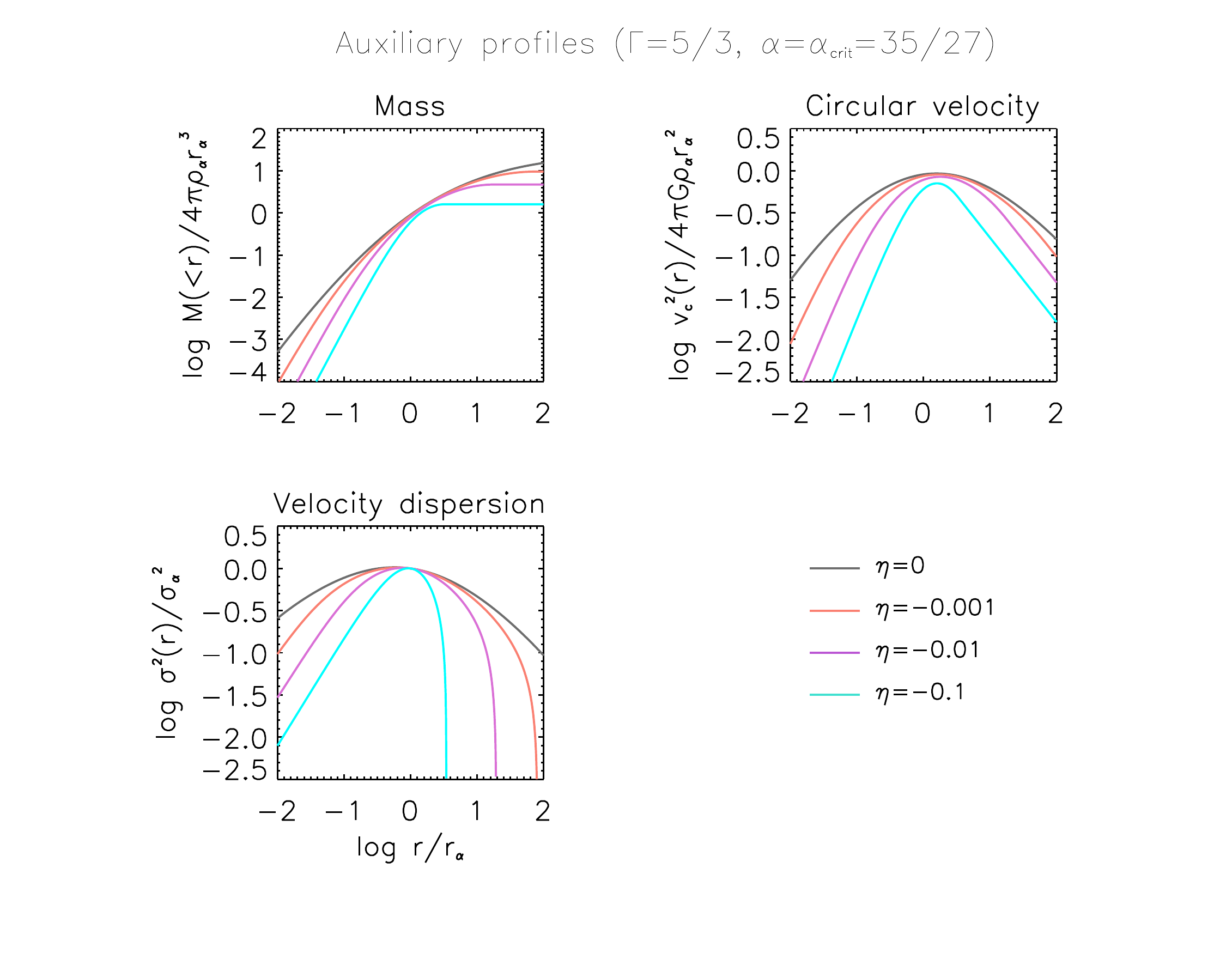}
\caption{Profiles of mass (top left panel), circular velocity (top right panel) and velocity dispersion (bottom left panel) corresponding to a few of the density profiles illustrated in the previous Figure (EOS parameters $\Gamma=\frac{5}{3}$ and $\alpha=\alpha_{\rm crit}=\frac{35}{27}$), with $\eta=0$ (black), $-0.001$ (orange), $-0.01$ (magenta), $-0.1$ (cyan).}\label{fig|NMC_auxiliary}
\end{figure}

\begin{figure}
\centering
\includegraphics[width=0.8\textwidth]{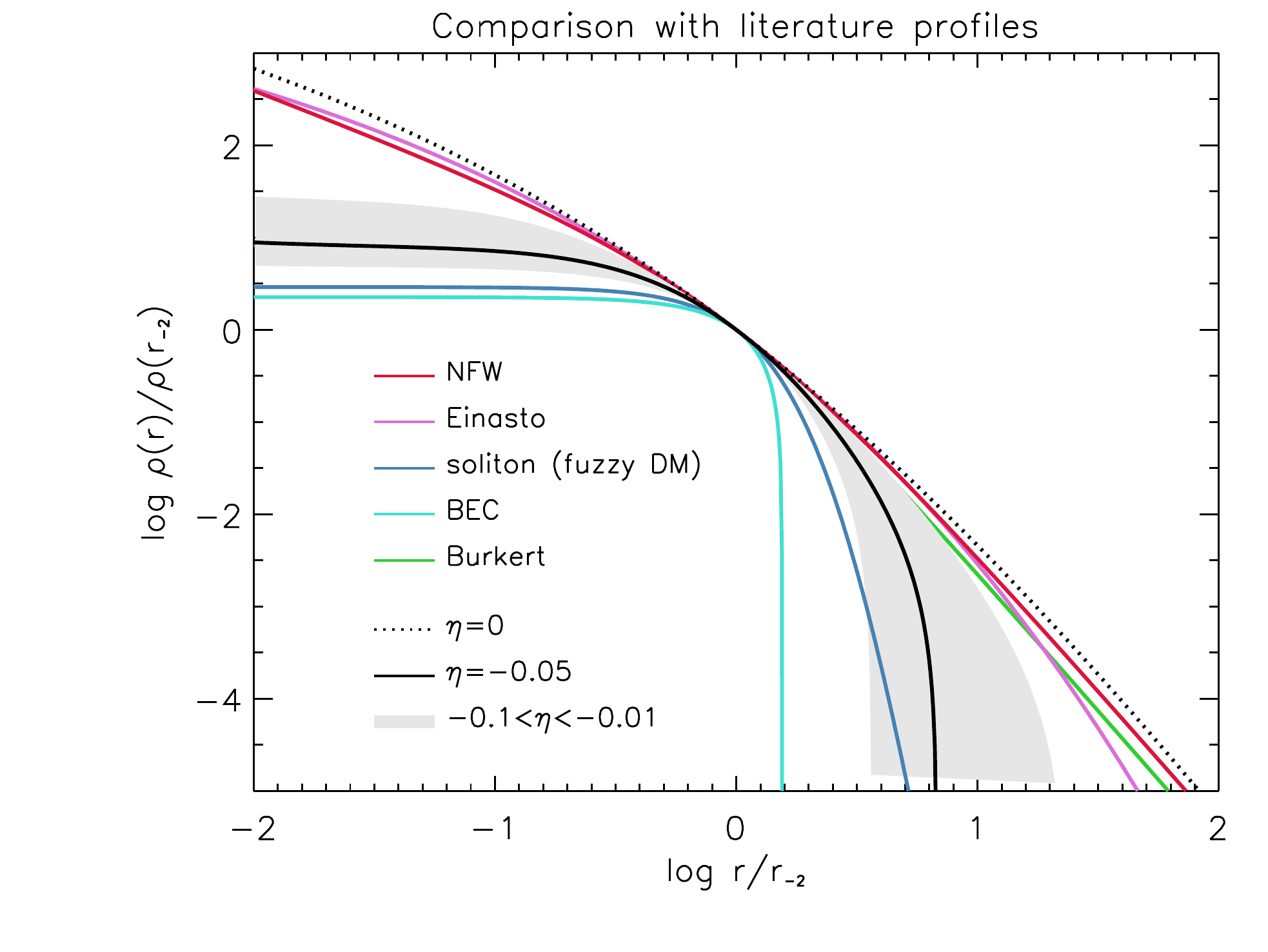}
\caption{Comparison of non-minimally coupled density distributions with classic literature profiles (see Sect.~\ref{sec|profcomp} for details): NFW (red line), Einasto (magenta line), soliton (blue line), interacting Bose-Einstein condensate (cyan line), and Burkert (green line). The dot black line refers to $\eta=0$, the solid black line is for $\eta=-0.05$ and the grey-shaded area illustrates the region covered by $\eta$ in the range from $-0.1$ (lower envelope) to $-0.01$ (upper envelope). The radial coordinate and the density profiles have been normalized to the radius $r_{-2}$ where the logarithmic density slope $\frac{{\rm d}\log\rho}{{\rm d}\log r}=-2$.}\label{fig|NMC_density_comp}
\end{figure}

\subsection{The fundamental equation and its solutions}\label{sec|fundeq}

It is convenient to introduce normalized variables $\bar r\equiv r/r_\alpha$, $\bar\rho\equiv \rho/\rho_\alpha$ and define the quantities $\kappa\equiv 4\pi G\, \rho_\alpha\,r_\alpha^2/\sigma_\alpha^2$ and $\eta\equiv \epsilon\, L^2/r_\alpha^2$. To understand the physical meaning of $\kappa$, one can choose $r_\alpha=r_{\rm max}$ to be the point at which the circular velocity $v_c^2(r)\equiv G\, M(<r)/r$ peaks at a value $v_c^2(r_{\rm max})=4\pi\rho(r_{\rm max})\, r_{\rm max}^2$, so that $\kappa = v_c^2(r_{\rm max})/\sigma^2(r_{\rm max})$ is seen to
compare the estimate $\sigma^2(r_{\rm max})$
for the random kinetic energy with that $v_c^2(r_{\rm max})$ for the gravitational potential.

Eliminating ${\rm d\Phi}/{\rm d}r$ from Eqs.~(\ref{eq|static}) and using the EOS Eq.~(\ref{eq|EOS}) yields the following fundamental equation for the density
\begin{equation}\label{eq|fundeq}
\bar\rho'' + \cfrac{(\Gamma-2)\,\cfrac{\bar\rho'^2}{\bar\rho}+
\cfrac{\alpha\,(2\Gamma-1)+2\,\Gamma}{\Gamma}\, \cfrac{\bar\rho'}{\bar r}+
\cfrac{\alpha\,(\alpha+1)}{\Gamma}\,\cfrac{\bar\rho}{\bar r^2}-2\,\eta\,\kappa\,\cfrac{\bar\rho^{2-\Gamma}\, \bar \rho'}{\Gamma\, \bar r^{\alpha+1}}+\kappa\,\cfrac{\bar \rho^{3-\Gamma}}{\Gamma\, \bar r^\alpha}}{{1-\eta\,\kappa\,\cfrac{\bar \rho^{2-\Gamma}}{\Gamma\,\bar r^\alpha}}}=0~,
\end{equation}
while the mass conservation constraint now reads $\mathcal{M}=\rho_\alpha\,r_\alpha^3\, f_{\mathcal{M}}$ in terms of the shape factor $f_{\mathcal{M}}=4\pi\,\int_0^{\infty}{\rm d}\bar r\, \bar r^2\, \bar \rho(\bar r)$. The solution space of such an equation is amazingly rich, and for $\Gamma=\frac{5}{3}$ and $\eta=0$ it has been quite extensively studied in the literature to describe the radial structure of standard $\Lambda$CDM halos (see Williams et al. 2004; Hansen 2004; Austin et al. 2005; Dehnen \& McLaughlin 2005; Lapi \& Cavaliere 2009a). In the following we provide the generalization with a generic $\Gamma$ and then with the addition of the non-minimal coupling.

To understand the general features of the solutions, it is convenient to look for powerlaw behaviors $\bar\rho\simeq \bar r^{-\gamma}$.
Substituting in the fundamental equation yields
\begin{equation}\label{eq|fundeqpowlaw}
\Gamma\,(\Gamma-1)\, \left(\gamma-\frac{\alpha}{\Gamma}\right)\,\left(\gamma-\frac{\alpha+1}{\Gamma-1}\right)-
\eta\,\kappa\,\frac{\gamma\,(\gamma-1)}{\bar r^{\gamma\,(2-\Gamma)+\alpha}} = -\cfrac{\kappa}{\bar r^{\gamma\,(2-\Gamma)+\alpha-2}}~.
\end{equation}
Focusing first on the minimally coupled case with $\eta=0$, it is evident that trivial powerlaw solutions with slope $\gamma=\gamma_\alpha\equiv \frac{2-\alpha}{2-\Gamma}$ are admitted, implying $\kappa=\kappa_{\rm PL}\equiv 2\,\frac{(\Gamma-\alpha)\,(\alpha+4-3\Gamma)}{(2-\Gamma)^2}$. These values are actually not physically acceptable at small and large radii since the gravitational force and mass would diverge, but provide the behavior of any solution at intermediate radii. In fact, to find general solutions it is numerically convenient to choose $r_\alpha$ as the reference radius where the logarithmic slope of the density is $-\gamma_\alpha$ and then integrate inward and outward. In terms of the normalized variables, this corresponds to set the boundary conditions as $\bar\rho(1)=1$ and $\bar\rho'(1)=-\gamma_\alpha$. When integrating inward of $r_\alpha$, it is found that the density profile features a physically acceptable behavior only for a specific value of $\kappa=\kappa_{\Gamma,\alpha}$, somewhat different from the $\kappa_{\rm PL}$ defined above. In particular, for such value $\kappa_{\Gamma,\alpha}$ the profile asymptotes for $\bar r\ll 1$ to $\bar\rho\propto \bar r^{-\gamma_0}$ with $\gamma_0\equiv \frac{\alpha}{\Gamma}$, consistently with the power counting in Eq.~(\ref{eq|fundeqpowlaw}). For $\kappa>\kappa_{\Gamma,\alpha}$ the profile has wiggles (change of sign in the second derivative) and steepens toward the center to imply a diverging gravitational force, while for $\kappa<\kappa_{\Gamma,\alpha}$ it develops a central hole. The optimal value $\kappa_{\Gamma,\alpha}$ can be found by a shooting technique (i.e., automatically solving inward the differential equation with different slopes $-\gamma_\alpha$ at $r_\alpha$ until the desired, physical inner asymptotic behavior is found), as represented in Fig.~\ref{fig|NMC_shooting}.

Fig.~\ref{fig|NMC_density_alpha} illustrates the resulting full density profiles, for three different values of $\alpha$. The outer behavior of the profile turns out to be physical only for $\alpha\leq \alpha_{\rm crit}\equiv \frac{\Gamma(5\Gamma-6)}{3\Gamma-2}$. For $\alpha>\alpha_{\rm crit}$ the outer slope is too flat, to imply a diverging mass.
For $\alpha<\alpha_{\rm crit}$ the solution $\bar\rho\propto \bar r^{-\gamma_\infty}$ attains a slope $\gamma_\infty\equiv \frac{\alpha+1}{\Gamma-1}$ at a finite large radius before an outer cutoff. For $\alpha=\alpha_{\rm crit}$ the cutoff is pushed to infinity and the slope $\gamma_{\infty,\rm crit}\equiv \frac{5\Gamma+2}{3\Gamma-2}$ is attained only asymptotically for $\bar r\rightarrow \infty$, consistently with the power-counting from  Eq.~(\ref{eq|fundeqpowlaw}); correspondingly, the intermediate and inner slopes read $\gamma_{\alpha,\rm crit}\equiv \frac{5\Gamma-2}{3\Gamma-2}$ and $\gamma_{0,\rm crit}\equiv \frac{5\Gamma-6}{3\Gamma-2}$, respectively. We illustrate these different behaviors in Fig.~\ref{fig|NMC_density_alpha} for the fiducial value $\Gamma=\frac{5}{3}$. In such a case, the powerlaw solutions have $\gamma_\alpha=6-3\,\alpha$ and $\kappa_{\rm PL}=6\,(5-3\,\alpha)\,(\alpha-1)$, while the physical solutions feature an inner slope $\gamma_0=\frac{3\,\alpha}{5}$ and an outer slope $\gamma_\infty=\frac{3\,(1+\alpha)}{2}$; the values of $\kappa_{\alpha,\Gamma}$ are reported as a function of $\alpha$ in the inset of Fig.~\ref{fig|NMC_density_alpha}. For $\alpha=\alpha_{\rm crit}=\frac{35}{27}$ one gets $\gamma_{0,\rm crit}\equiv \frac{7}{9}$, $\gamma_{\alpha,\rm crit}=\frac{19}{9}$, $\gamma_{\infty,\rm crit}= \frac{31}{9}$, and $\kappa_{\alpha,\Gamma}=\frac{200}{81}\approx 2.5$.

When including the non-minimal coupling $\eta\equiv \epsilon\,\frac{L^2}{r_\alpha^2}$, the solution space changes appreciably. First of all, $\epsilon$ (hence $\eta$) must be negative, otherwise the inner profile diverges at a finite radius. Then for any negative value of $\eta$, there is again an optimal value of $\kappa=\kappa_{\Gamma,\alpha,\eta}$ such that the inner profile is physical, with limiting central slope $\gamma_0=0$, i.e. a core. This is again in accordance with the power counting in  Eq.~(\ref{eq|fundeqpowlaw}) since now the second term on the l.h.s. dominates the behavior for small $\bar r\ll 1$. Other solutions with $\kappa$ smaller or larger than $\kappa_{\Gamma,\alpha,\eta}$ are not physically acceptable since they have non-monotonic behaviors with the density first flattening and then steepening toward a central slope $\gamma_0=1$, or they develop a central hole. As for the outer behavior, the profile has a cutoff at a finite radius $\mathcal{R}$ setting the effective halo boundary, which is smaller for more negative values of $\eta$. We illustrate the physically acceptable profiles for different $\eta$ and the related values of the constant $\kappa_{\Gamma,\alpha,\eta}$ in Fig.~\ref{fig|NMC_density_eta}. The corresponding distributions of mass, circular velocity and velocity dispersion for a few values of the non-minimal coupling $\eta$ are also illustrated in Fig.~\ref{fig|NMC_auxiliary}.

\section{Comparison with literature profiles}\label{sec|profcomp}

We now compare the shape of our physical solutions to some classic literature density profiles, characterized by different analytic expressions and numbers of parameters, that are commonly adopted to fit simulations and/or observations (see also Freundlich et al. 2020b, their Fig. 15 for a comprehensive account). To this purpose, it is convenient to use a radial coordinate $\hat r\equiv r/r_{-2}$ normalized to the radius $r_{-2}$ where the logarithmic density slope $\frac{{\rm d}\log \rho}{{\rm d}\log r}=-2$, and to rescale the density profile $\hat \rho\equiv \rho/\rho(r_{-2})$ accordingly. Specifically, we will consider the following density profiles.

\begin{itemize}

  \item \underline{$\alpha\beta\gamma$ profiles}

  The $\alpha\beta\gamma$ profiles (Zhao 1996; see also Widrow 2000) feature the shape
  \begin{equation}\label{eq|gNFW}
  \hat\rho(\hat r)=\hat r^{-\tau}\, \left(\cfrac{1+w}{1+w\,\hat r^\omega}\right)^\xi~,
  \end{equation}
  where the three parameters $\tau$, $\omega$, $\xi$ describe respectively the central slope, the middle curvature and the outer decline of the density run, while $w\equiv - \frac{2-\tau}{(2-\tau-\omega\,\xi)}$. Familiar empirical profiles are recovered for specific values of the triplet $(\tau,\omega,\xi)$: e.g., Plummer's profile corresponds to $(0,2,5/2)$, Jaffe's to (2,1,2), and Hernquist's to $(1,1,3)$. The standard NFW (Navarro et al. 1996), which is classically used to fit $N-$body simulations in the $\Lambda$CDM model, is obtained for the parameter triple $(1,1,2)$; for a generalization with different inner slope, often referred to as gNFW, the parameters $(\tau,1,3-\tau)$ apply (Mamon et al. 2019; for a more complex cored version see also Read et al. 2016). It is worth mentioning that recently Freundlich et al. (2020b) have considered a $\alpha\beta\gamma$ profile (referred also as Zhao-Dekel model) with parameters $(\tau,\frac{1}{2},7-2\tau)$, that can provide good fits to the density profiles from both $N-$body, DM-only and hydro simulations including baryonic effects.

  \item \underline{Sersic-Einasto profile}

  The Sersic-Einasto profile (see An \& Zhao 2013) is defined as
  \begin{equation}\label{eq|Einasto}
  \hat\rho(\hat r)=\hat r^{-\tau}\, e^{-u\,(\hat r^\omega-1)}~,
  \end{equation}
  where $\tau$ is the inner density slope, $\omega$ is a shape parameter and $u\equiv \frac{(2-\tau)}{\omega}$. The classic cored Einasto shape (Sersic 1963; Einasto 1965; Prugniel et al. 1997; Graham et al. 2006; also Lazar et al. 2020 for a more complex analytical expression) is recovered for $\tau=0$. $N-$body simulations in the standard $\Lambda$CDM model are usually well described by the parameter values $\omega\approx 0.15-0.2$ and $\tau\approx 0$ (although values $\tau\lesssim 0.8$ are not ruled out given the resolution of current simulations).

  \item \underline{Soliton profile}

  The soliton profile features the shape (see Schive et al. 2014a,b)
  \begin{equation}\label{eq|soliton}
  \hat\rho(\hat r)=\left(\cfrac{1+\omega/\hat r_c^2}{1+\omega\,\hat r^2/\hat r_c^2}\right)^8~,
  \end{equation}
  where the normalized core radius reads $\hat r_c = \sqrt{7\,\omega}$. In numerical simulations of non-interacting Bose-Einstein condensate DM (alias fuzzy DM), halos are described by a combination of this solitonic profile with $\omega\approx 0.091$ in the inner region, and of a NFW profile in the outskirts.

  \item \underline{BEC profile}

  The profile followed by an interacting Bose-Einstein condensate (BEC; see Bohmer \& Harko 2007; Harko 2014) in the Thomas-Fermi limit reads
  \begin{equation}\label{eq|BEC}
  \hat\rho(\hat r)= \cfrac{1}{\hat r}\,\cfrac{\sin(\pi\,\hat r/\hat R)}{\sin(\pi/\hat R)}~,
  \end{equation}
  where the normalized halo boundary is defined by the equality $\pi/\hat R + \tan(\pi/\hat R)=0$. Incidentally, note that such a profile actually corresponds to the solution of our Eq.~(\ref{eq|fundeqpowlaw}) for $\Gamma=2$ and $\alpha=\eta=0$.

  \item \underline{Burkert profile}

  The Burkert profile can be written as (see Burkert 1995; Salucci \& Burkert 2000)
  \begin{equation}\label{eq|Burkert}
  \hat\rho(\hat r)=\cfrac{(1+\hat r_c)\,(1+\hat r_c^2)}{(\hat r+\hat r_c)\,(\hat r^2+\hat r_c^2)}~,
  \end{equation}
  where $\hat r_c$ is the normalized core radius, defined by the nonlinear algebraic equation $2\,\hat r_c^3+\hat r_c^2-1=0$. This is a cored profile generally exploited to fit observations of (dwarf) spiral galaxies.

\end{itemize}

In Fig.~\ref{fig|NMC_density_comp} we compare the shape of our  solutions (for definiteness $\Gamma=\frac{5}{3}$ and $\alpha=\alpha_{\rm crit}=\frac{35}{27}$ are adopted) to a few of the above profiles. First it is evident that our density run for $\eta=0$ describes quite well the NFW and Einasto profiles commonly used to fit $N-$body simulations in the standard $\Lambda$CDM cosmology. For $-0.1<\eta<-0.01$ our solutions develop a core and the shape out to few/several $r_{-2}$ is remarkably close to the Burkert profile, commonly exploited to fit observations of dwarf galaxies. At larger distances there is progressive deviation from the Burkert profile, since the latter has been designed to have a limiting slope close to the NFW one, while our profiles get truncated at a finite radius $\mathcal{R}$. However, for an appreciable range of $\eta$ values this is not a concern since the truncation occurs at radii much beyond $r_{-2}$ that are scantily (if at all) probed by observations (see also Sect.~\ref{sec|datacomp}). We stress that it is a remarkable property of the non-minimally coupled solutions to reproduce the Burkert shape not only in the inner region, but also over an extended radial range outward of $r_{-2}$. For comparison, other models such as those based on Bose-Einstein condensate DM predict a cored density profile, but the deviation from the Burkert shape are appreciable, in terms of a prominent cutoff or steep decline, just outside the core. Note that our solutions for $\eta=-0.05$ can be reasonably described by a $\alpha\beta\gamma$ model (cf. Eq.~\ref{eq|gNFW}) with parameters $(\tau,\omega,\xi)=(0,\frac{3}{2},\frac{5}{2})$, or by a Sersic-Einasto model (cf. Eq.~\ref{eq|Einasto})  with parameters $(\tau,\omega)=(0,\frac{3}{4})$ up to several $r_{-2}$ before the final cutoff.

\begin{figure}
\centering
\includegraphics[width=0.8\textwidth]{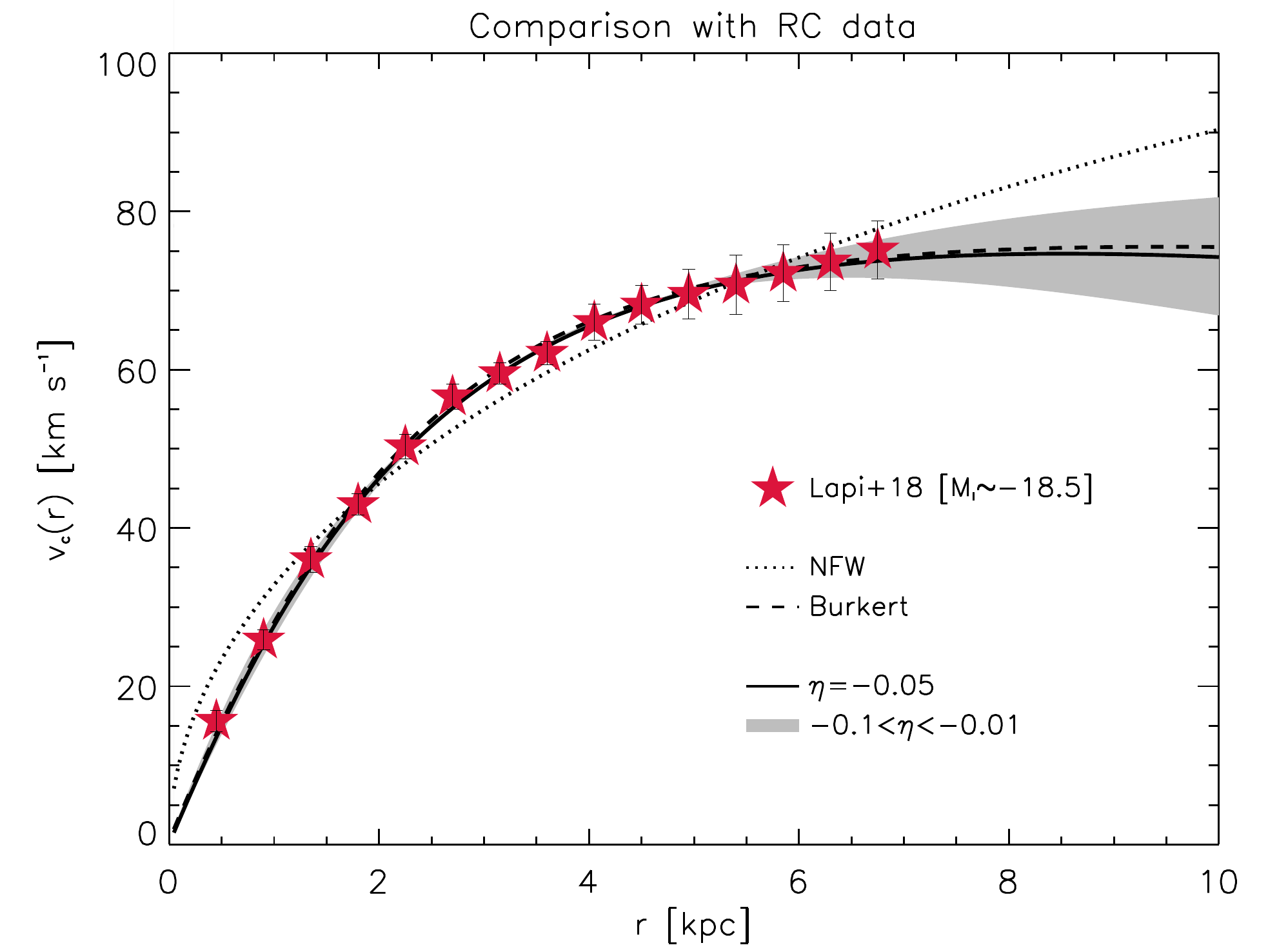}
\caption{Comparison of non-minimally coupled halo mass distributions with observed dwarf galaxy rotation curves. Data points (red stars; Lapi et al. 2018) refer to the co-added rotation curves of about $20$ dwarf galaxies with I-band magnitude $M_I\ga -18.5$, extracted from the original sample by Persic \& Salucci (1996). Solid line illustrates the fit via our physical solution with non-minimal coupling parameter $\eta\approx -0.05$, while the shaded grey area show the effect of changing $\eta$ in the range from $-0.1$ to $-0.01$. For comparison, dashed line illustrates the fit with the Burkert profile, and dotted line that with the NFW profile.}\label{fig|datacomp}
\end{figure}

\section{Comparison with measured rotation curves}\label{sec|datacomp}

We now compare the non-minimally coupled mass distributions with observed galaxy rotation curves. To avoid dealing with baryons (which are not included in our treatment), we focus on dwarf, strongly DM-dominated galaxies. Specifically, we exploit the co-added rotation curve built by Lapi et al. (2018) based on the high-quality measurements of about $20$ dwarf galaxies with $I-$band magnitude $M_I\ga -18.5$ from the original sample by Persic \& Salucci (1996). The sample features an average disk scale-length $R_e\la 1.5$ kpc, disc mass $M_\star\la 10^{9}\, M_\odot$ and halo mass $\mathcal{M}\la 10^{11}\, M_\odot$. The co-added rotation curve is well-measured out to a galactocentric distance of about $7$ kpc.
We then use a Levenberg-Marquardt least-squares minimization routine to fit the measured rotation curve with the NFW and Burkert profile, and with our
physical solutions for different values of the non-minimal coupling parameter $\eta$; the outcomes are illustrated in Fig.~\ref{fig|datacomp}.

As it is well known, the NFW fit (equivalent to our solution with $\eta=0$) struggles to fit the measured dwarf galaxy rotation curves, yielding a reduced $\chi^2\approx 4$. On the other hand, the Burkert profile performs much better, providing a good fit with a reduced $\chi^2\approx 0.58$. Our non-minimally coupled solution for $\eta\approx -0.05$ provides a fit of quality comparable to the Burkert one, yielding practically the same reduced $\chi^2\approx 0.61$. The current data are compatible within $3\sigma$ with any value of $\eta$ in the range from $-0.1$ to $-0.01$. Accurate determinations of the co-added rotation curve out to $10$ kpc or beyond would be necessary to determine the non-minimal coupling parameter $\eta$ to a good level of precision. Interestingly, one can also see from Fig.~\ref{fig|datacomp} that there is a tendency to favor values of $\eta$ slightly less negative than $-0.05$, which is actually what is to be expected in dwarf galaxies with halo masses $\mathcal{M}\la 10^{11}\, M_\odot$ on the basis of universal scaling arguments (see Sect.~\ref{sec|scaling} and in particular Eq.~\ref{eq|etamass}).

From the fit with $\eta\approx -0.05$ (or equivalently from the Burkert one) we derive a best fit value of the core radius around $r_0\approx 2.9\pm 0.1$ kpc, which is about twice the average disk-scale length of the systems in the considered sample. This is remarkably consistent with the empirical, yet still puzzling relationship between core radius and disc scale-length as determined by Donato et al. (2004).

A remark is in order concerning the good agreement of our simple non-minimally coupled DM model with the observed rotation curve. On the one hand, Bruneton et al. (2009) have shown that choosing an appropriate yet rather complex shape (see their Eqs. 2.6-2.7) of the coupling $h_{\mu\nu}$ characterizing the interaction term in the action of Eq.~(\ref{eq|jordan}) can lead to a MONDian phenomenology on galactic scales, that is well known to reproduce rotation curves. However, our model is based on a much simpler coupling $h_{\mu\nu}\propto \nabla_\mu\varphi\nabla_\nu\varphi$, and hence does not lead exactly to MONDian dynamics in its Newtonian limit; from this perspective, the rather good performance of our simple non-minimally coupled model in reproducing the measured rotation curves is valuable and far from trivial.

\begin{figure}
\centering
\includegraphics[width=0.8\textwidth]{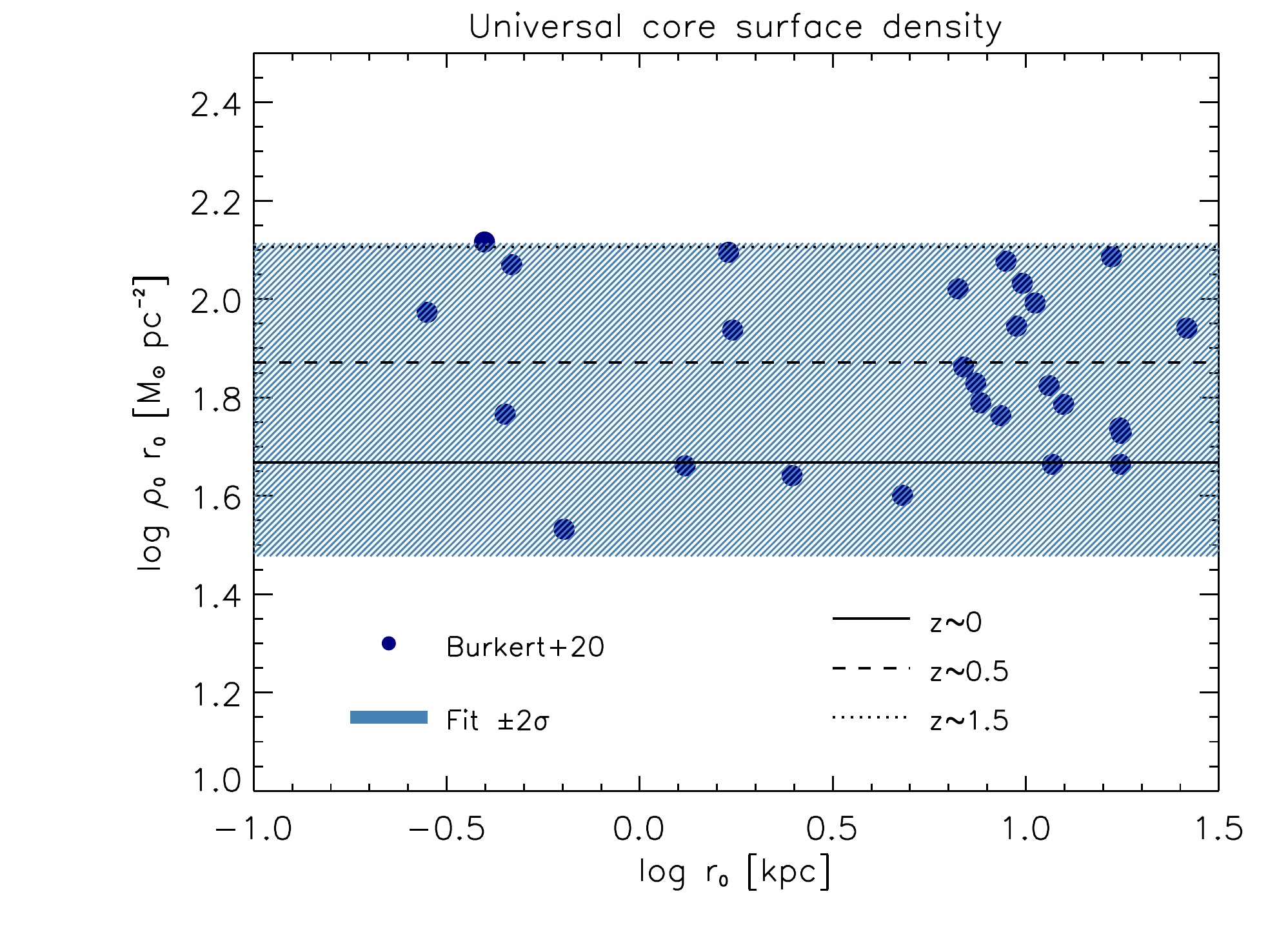}
\caption{Core surface density $\rho_0\times r_0$ as a function of the core radius $r_0$. Data points from Burkert (2020) are well fitted by the universal value $75_{-45}^{+55}\, M_\odot$ pc$^{-2}$ (blue shaded area). Our prediction for non-minimally coupled DM halos is illustrated by the black lines for three different halo formation redshifts $z\approx 0$ (solid), $0.5$ (dashed), and $1.5$ (dotted).}\label{fig|NMC_scaling}
\end{figure}

\section{Universal core surface density}\label{sec|scaling}

As mentioned in Sect.~\ref{sec|intro}, it has been well established observationally (see Salucci \& Burkert 2000; Burkert 2015) that, at least for dwarf galaxies with halo masses $\mathcal{M}\la 10^{11}\, M_\odot$, the product of the core density and core radius (i.e., a sort of core surface density) is an approximately universal constant with values $\rho_0\, r_0\approx 75_{-45}^{+55}\, M_\odot$ pc$^{-2}$ among different galaxies, see Fig.~\ref{fig|NMC_scaling}. This somewhat unexpected property poses a serious challenge to any theoretical model of core formation (e.g., Deng et al. 2018; Burkert 2020). We now aim to show that non-minimally coupled DM halos are instead consistent with such a remarkable scaling law.

First, adopting for definiteness $\Gamma=\frac{5}{3}$ and $\alpha=\alpha_{\rm crit}=\frac{35}{27}$, we compute the physical solutions of Eq.~(\ref{eq|fundeq}) subject to the boundary conditions $\rho(r_\alpha)=\rho_\alpha$ and $\rho'(r_\alpha)=-\gamma_\alpha\, \frac{\rho_\alpha}{r_\alpha}$, for several values of the non-minimal coupling parameter $\eta$. From the so obtained normalized profiles we fit as a function of $|\eta|=\frac{L^2}{r_\alpha^2}$ in the range $-0.1\lesssim\eta\lesssim -0.01$ the following relations involving the normalized boundary radius $\mathcal{R}$, the core radius $r_0$, the core density $\rho_0$ and the mass shape factor $f_{\mathcal{M}}$ defined below Eq.~(\ref{eq|fundeq}):
\begin{equation}
\frac{\mathcal{R}}{r_\alpha}\simeq 0.8\, |\eta|^{-0.7}~,~~~~~~
f_{\mathcal{M}}\simeq 9.1\, |\eta|^{-0.4}~,~~~~~~\frac{\rho_0}{\rho_\alpha}\simeq 0.9\, |\eta|^{-0.8}~,~~~~~~\frac{r_0}{r_\alpha}\simeq 1.6\, |\eta|^{0.5}~.
\end{equation}

We then combine the above scaling with the expression for the total mass $\mathcal{M}=f_{\rm \mathcal{M}}\, \rho_\alpha\,r_\alpha^3$ and with the definition of the virial radius $\mathcal{R}=(3\, \mathcal{M}/4\pi\Delta_{\rm vir}\, \rho_{\rm c}\, E_z)^{1/3}\approx 120\,E_z^{-1/3}\, (\mathcal{M}/10^{11}$ $M_\odot)^{1/3}$; here $\rho_{\rm c}\approx 2.8\times 10^{11}\, h^2\,M_\odot$ Mpc$^{-3}$ is the critical density, $E_z\equiv \Omega_M\,(1+z)^3+\Omega_\Lambda$ takes into account the formation redshift $z$ of the halo, and $\Delta_{\rm vir}$ is the nonlinear threshold for virialization, with values around $100$ at $z\approx 0$ and increasing toward $180$ for $z\ga 1$. We eventually derive
\begin{equation}\label{eq|scaling}
r_0\simeq 1.6\, L~,~~~~~~~r_\alpha\simeq 1.1\, L^{0.6}\, \mathcal{R}^{0.4}~~~~~~\rho_0\simeq 0.3\,\Delta_{\rm vir}\,\rho_{\rm c}\,E_z\, \left(\frac{\mathcal{R}}{L}\right)^{2.1}=\rho_\alpha\,\left(\frac{\mathcal{R}}{L}\right)^{0.6}~;
\end{equation}
remarkably, the core radius $r_0$ turns out to be proportional, with a coefficient of order $1$, to the non-minimal coupling length-scale $L$.

To proceed further we need a relation between the core radius $r_0$ (or $L$) and the halo mass $\mathcal{M}$; this is thought not to be of fundamental nature but rather to stem from two other relationships involving the baryonic
mass: (i) the relation between the stellar (disc) mass and the halo mass (e.g., Moster et al. 2013), which is known to be originated by baryonic processes related to galaxy formation; (ii) the relation between the core radius $r_0$ and the disk scalelength (in turn related to the stellar mass; e.g., Donato et al 2004), which is instead still not completely understood. In the present pilot study we are not including the baryonic component and we cannot infer the $r_0-\mathcal{M}$ relation from first principles; thus we will adopt the outcome $r_0\approx 4.5\, (\mathcal{M}/10^{11}\, M_\odot)^{0.6}$ kpc from the dynamical modeling study by Salucci et al. (2007), and see what this implies for the core surface density.
Specifically, from Eqs.~(\ref{eq|scaling}) we obtain
\begin{equation}
\Sigma_0\equiv \rho_0\times r_0 \approx 50\, \left(\frac{\Delta_{\rm vir}}{100}\right)\,E_z^{0.3}\,M_\odot\,{\rm pc^{-2}}~
\end{equation}
independent of the halo mass and/or core radius, and only weakly dependent on formation redshift. In Fig.~\ref{fig|NMC_scaling} we report the above for three values of the formation redshift $z\approx 0$, $0.5$, $1.5$ finding it remarkably consistent with the average observed relation and its scatter.
We stress the extreme relevance of this finding: the universality of the core surface density has proven to be extremely challenging for alternative DM models, even for those that are barely consistent with the $r_0-\mathcal{M}$ relation we have assumed. For example, as pointed out by Burkert (2020) fuzzy DM can reproduce the $r_0-\mathcal{M}$ relation, albeit with some (uncertain) hypothesis on core formation redshift. However, such a model is considerably out of track as to the core surface density scaling, since it robustly predicts $\rho_0\propto r_0^{-4}$. The same issue concerns many other DM models inspired by particle physics, as extensively discussed, e.g., by Deng et al. (2018).

As an aside, from Eqs.~(\ref{fig|NMC_scaling}) and the adopted $r_0-\mathcal{M}$ relation we can also derive other three interesting scaling laws. First, the dependence of $\eta$ on halo mass reads
\begin{equation}\label{eq|etamass}
|\eta|\approx 0.04\, \left(\frac{\mathcal{M}}{10^{11}\, M_\odot}\right)^{0.35}\,E_z^{0.3}~;
\end{equation}
this confirms that values of the non-minimal coupling in the range $\eta=-0.1$ to $-0.01$ cover the typical mass range of dwarf galaxies. We stress that this dependence of  $\eta$ on halo mass/formation redshift will induce slightly different shapes in the profiles, implying a weak violation of self-similarity. Though challenging, it will be interesting to look for such behaviors in real data (see Sect.~\ref{sec|datacomp}).
Second, it may be interesting to derive the dependence on the coupling, hence on mass, of the inner logarithmic slope $\gamma_{0.1}$ measured at a reference radius of $r\approx 0.1\, r_{-2}\sim$ a few percent of $\mathcal{R}$. We get the scaling $|\gamma_{0.1}|\simeq 0.035\,|\eta|^{-0.55}$ that after Eq.~(\ref{eq|etamass}) translates into a mass-dependence
\begin{equation}
|\gamma_{0.1}|\simeq  0.2\, \left(\frac{\mathcal{M}}{10^{11}\, M_\odot}\right)^{-0.2}\,E_z^{-0.16}~.
\end{equation}
Thus there is a slight tendency for less massive halos to have flatter profile at a fixed radius in the inner region (note that asymptotically at the center all the non-minimally coupled DM density profiles are flat). In a future work, it would be interesting to investigate how such a scaling is altered by the presence of baryons in halos of different masses, and how the outcome will compare with the results from $\Lambda$CDM hydrodynamical simulations including feedback effects (e.g., Tollet et al. 2016; Freundlich et al. 2020b), that show a non-trivial mass dependence for the halo inner shape. Third, we can compute the halo concentration as $c_\alpha\equiv \frac{\mathcal{R}}{r_\alpha}$ (using $\frac{\mathcal{R}}{r_{-2}}$ yields similar result), which turns out to be
\begin{equation}
c_\alpha\simeq  10\, \left(\frac{\mathcal{M}}{10^{11}\, M_\odot}\right)^{-0.15}\,E_z^{-0.2}~,
\end{equation}
in broad agreement, and actually slightly smaller than the outcome of $N-$body, DM-only simulations in the standard $\Lambda$CDM cosmology (e.g., Bullock et al. 2001; Macci\'o et al. 2007).

\section{Summary}\label{sec|summary}

We have investigated self-gravitating equilibria of halos constituted by dark matter (DM) non-minimally coupled to gravity. A non-minimal coupling may be present in modified gravity theories or it might be dynamically generated when the averaging/coherence length $L$ associated to a fluid description of the DM collective behavior is comparable to the local curvature scale. We have theoretically motivated a form of such a coupling that in the Newtonian limit amounts to a modification of the Poisson equation by a term $L^2\,\nabla^2\rho$ proportional to the Laplacian of the DM density $\rho$ itself (see Sect.~\ref{sec|NMC-DM}). We have further adopted an effective power-law equation of state $p\propto \rho^{\Gamma}\, r^\alpha$ relating the DM dynamical pressure $p$ to density $\rho$ and radius $r$, as expected by phase-space density stratification during the gravitational assembly of halos in a cosmological context (see Sect.~\ref{sec|EOS}). In absence of the non-minimal coupling, we have confirmed previous findings that the DM density run $\rho(r)$ features a central density cusp and an overall shape mirroring the outcomes of $N-$body simulations in the standard $\Lambda$CDM cosmology, as described by the classic NFW or Einasto profiles (see Sect.~\ref{sec|fundeq}).

We have remarkably found that, when the non-minimal coupling is switched on, it causes the DM density profile to develop an inner core and a shape closely following, out to several core scale radii, the Burkert profile (see Sect.~\ref{sec|profcomp}). In addition, we have highlighted that our non-minimally coupled solutions can fit, with an accuracy comparable to the Burkert profile, the co-added rotation curve of DM-dominated dwarf galaxies (see Sect.~\ref{sec|datacomp}).
Finally, we have shown that non-minimally coupled DM halos are consistent with the observed scaling relation between the core radius $r_0$ and the core density $\rho_0$ in terms of an universal core surface density $\rho_0\times r_0$ among different galaxies, that has proven to be challenging for many other DM models (see Sect.~\ref{sec|scaling}).

A future development of this work will involve the study of the DM density profile in presence of baryons. In fact, the non-minimal coupling to gravity constitutes a natural and effective way to tightly link the DM and baryon properties. On the one hand, this could help to understand puzzling scaling relationships between the DM and the baryonic component, and to characterise the physical processes underlying the emergence of the non-minimal coupling length-scale $L$. On the other hand, this will allow us to probe the effectiveness of our solutions in fitting the measured rotation curves of normal, and not only dwarf, rotation-dominated galaxies. In parallel, we plan to extend the static investigation pursued in the present paper to time-dependent conditions, by implementing the non-minimal coupling inside a full $N-$body numerical simulation. On more general grounds, it would be worth to explore the effect of non-minimal coupling on large, cosmological scales, especially in connection with the dark energy phenomenology.

In conclusion, we have proposed that a non-minimal coupling between matter and gravity could constitute a crucial ingredient toward an improved description of realistic DM structures in a cosmological framework. We very much hope this novel perspective will contribute to shed light on some of the remaining mysteries concerning the DM component in cosmic structures.

\acknowledgments

We thank the referee for helpful comments and suggestions. We acknowledge D. Bettoni, L. Danese, M. Nori and P. Salucci for stimulating discussions and critical reading. This work has been partially supported by PRIN MIUR 2017 prot. 2017-3ML3WW, “Opening the ALMA window on the cosmic evolution of gas, stars and supermassive black holes.” A.L. has taken advantage of the MIUR grant “Finanziamento annuale individuale attivit\'a base di ricerca” and of the EU H2020-MSCA-ITN-2019 Project 860744 “BiD4BEST: Big Data applications for Black hole Evolution STudies.” S.L. acknowledges funding from the grant PRIN MIUR 2017 prot. 2017-MB8AEZ.

\newpage

\appendix

\section{Anisotropic conditions}\label{sec|anisotropy}

In this Appendix we discuss the self-gravitating equilibria of non-minimally coupled DM halos when anisotropic conditions apply. These can be included in our treatment by modifying the second of Eqs.~(\ref{eq|static}) as
\begin{equation}
\cfrac{1}{\rho}\,\cfrac{{\rm d}p}{{\rm d}r} + 2\beta\, \cfrac{\sigma_r^2}{r}= - \cfrac{{\rm d}\Phi}{{\rm d}r}~,
\end{equation}
where $\beta\equiv 1-\frac{\sigma_\theta^2}{\sigma_r^2}$ is the Binney (1978) anisotropy parameter in terms of the tangential and radial velocity dispersions $\sigma_\theta$ and $\sigma_r$, respectively. $N-$body simulations suggest $\beta(r)$ to increase from central values $\beta_0\lesssim 0$, meaning near isotropy, to outer values $\beta\gtrsim 0.5$, meaning progressive prevalence of radial motions. This overall trend can be physically understood in terms of efficient dynamical relaxation processes toward the inner regions, that tend to enforce closely isotropic conditions, while in the outskirts the infall energy of accreting matter is more easily converted by phase mixing into radial random motions (see Lapi et al. 2011 for details). Specifically, simulations  suggest the effective linear expression (e.g., Hansen \& Moore 2006)
\begin{equation}\label{eq|betaprof}
\beta(r)\simeq \beta_0+\beta_1\, [\gamma(r)-\gamma_0] =\beta_0-\beta_1\gamma_0-\beta_1\, r\, \cfrac{\rho'}{\rho}~,
\end{equation}
in terms of the logarithmic density slope $\gamma(r)\equiv -\frac{{\rm d}\log \rho}{{\rm d}\log r}$,
with $\gamma_0\equiv \gamma(0)$ being a value yet to be determined, $\beta_0\lesssim 0$ and $\beta_1\approx 0.2$.

Adopting Eq.~(\ref{eq|betaprof}) and following the same derivation of the main text, the fundamental Eq.~(\ref{eq|fundeq}) now reads
\begin{equation}\label{eq|fundeqanis}
\begin{aligned}
&\left[1-\eta\,\kappa\,\cfrac{\bar \rho^{2-\Gamma}}{(\Gamma-2\beta_1)\,\bar r^\alpha}\right]\,\bar\rho'' + (\Gamma-2)\,\cfrac{\bar\rho'^2}{\bar\rho}+\\
\\
&+ \cfrac{\alpha\,(2\Gamma-1)+2\,\Gamma+2(\Gamma-1)\beta_0-2[\alpha+2+(\Gamma-1)\gamma_0]\beta_1}{\Gamma-2\beta_1}\, \cfrac{\bar\rho'}{\bar r}+\\
\\
&+ \cfrac{[\alpha+1]\,[\alpha+2(\beta_0-\beta_1\gamma_0)]}{\Gamma-2\beta_1}\,\cfrac{\bar\rho}{\bar r^2}
-2\,\eta\,\kappa\,\cfrac{\bar\rho^{2-\Gamma}\, \bar \rho'}{(\Gamma-2\beta_1)\, \bar r^{\alpha+1}}+\kappa\,\cfrac{\bar \rho^{3-\Gamma}}{(\Gamma-2\beta_1)\, \bar r^\alpha}=0~,
\end{aligned}
\end{equation}
Looking for powerlaw behaviors $\bar\rho\simeq \bar r^{-\gamma}$ one obtains
\begin{equation}\label{eq|fundeqpowlawanis}
\Gamma\,(\Gamma-1)\, \left[\gamma-\frac{\alpha+2\beta_0}{\Gamma}-2\frac{\beta_1}{\Gamma}\,(\gamma-\gamma_0)\right]\,\left[\gamma-\frac{\alpha+1}{\Gamma-1}\right]-
\eta\,\kappa\,\frac{\gamma\,(\gamma-1)}{\bar r^{\gamma\,(2-\Gamma)+\alpha}} = -\cfrac{\kappa}{\bar r^{\gamma\,(2-\Gamma)+\alpha-2}}~,
\end{equation}
which, remarkably, allows to self-consistently determine the central slope $\gamma_0$.

In fact, for minimally coupled halos ($\eta=0$), the anisotropic solutions feature a modified inner slope $\gamma_0 = \frac{\alpha+2\beta_0}{\Gamma}$
with respect to the isotropic case, while retaining the same slopes at intermediate radii $\gamma_\alpha=\frac{2-\alpha}{2-\Gamma}$ and in the outer region $\gamma_\infty = \frac{\alpha+1}{\Gamma-1}$; as mentioned above, $\beta_0\approx 0$ so that the changes are minor (if any, $\beta_0\lesssim 0$ so that the inner profile is flattened a bit).
In addition, the critical solution is characterized by a value $\alpha_{\rm crit}=\frac{\Gamma\,(5\Gamma-6)+2\beta_0\,(\Gamma-1)\,(\Gamma-2)}{3\Gamma-2}$; in particular, $\alpha_{\rm crit}=\frac{35-4\beta_0}{27}$ holds for for $\Gamma=\frac{5}{3}$. The corresponding inner, intermediate and outer slopes read $\gamma_{0,\rm crit}=\frac{5\Gamma-6+2\beta_0\Gamma}{3\Gamma-2}$,  $\gamma_{\alpha,\rm crit}=\frac{5\Gamma-2+2\beta_0\,(\Gamma-1)}{3\Gamma-2}$, and $\gamma_{\infty,\rm crit}=\frac{5\Gamma+2+2\beta_0\,(\Gamma-2)}{3\Gamma-2}$, respectively.

\begin{figure}
\figurenum{A1}
\centering
\includegraphics[width=0.8\textwidth]{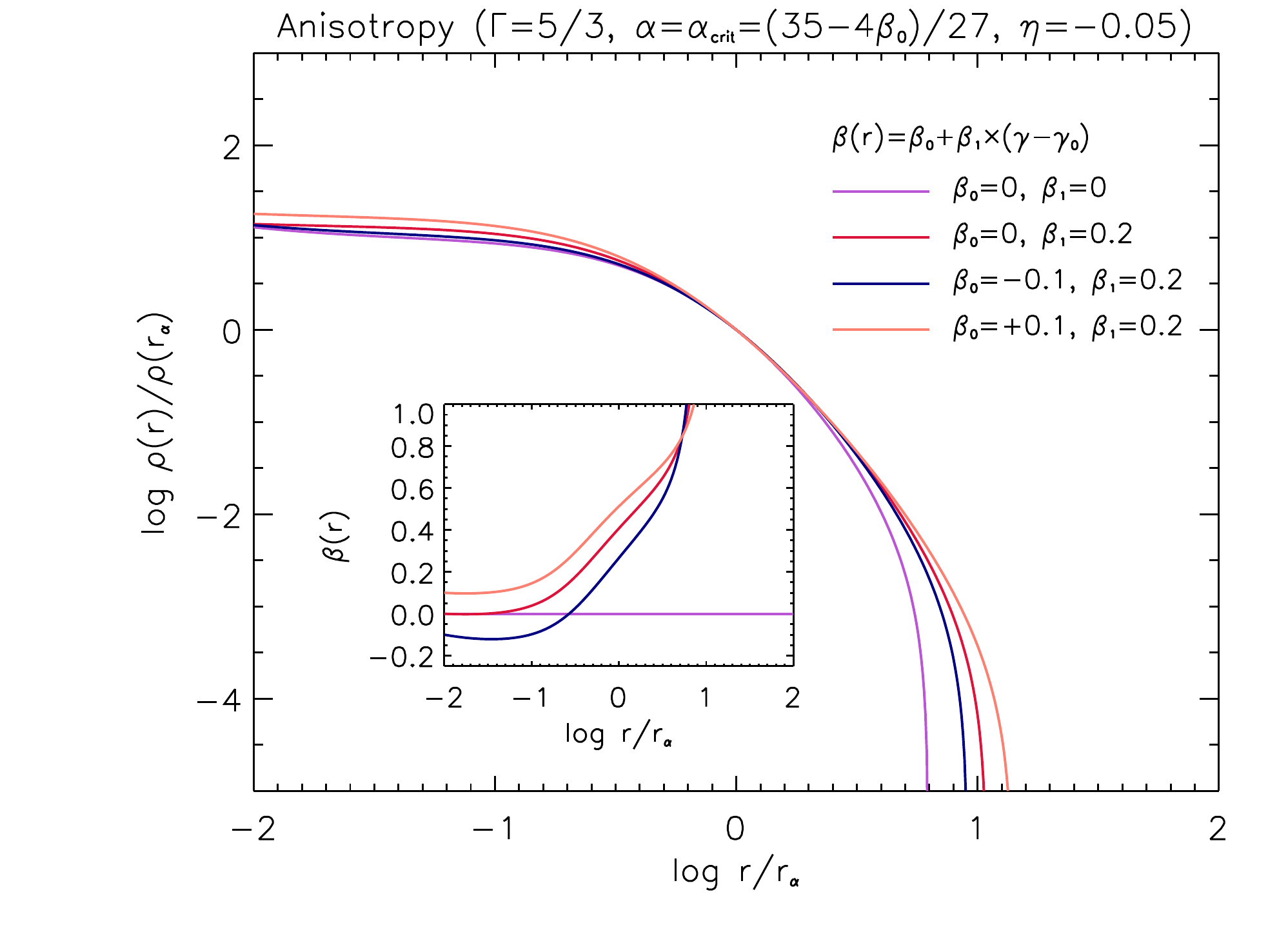}
\caption{Effects of realistic anisotropic conditions on the minimally-coupled density profile with $\eta=-0.05$; EOS parameters $\Gamma=\frac{5}{3}$ and $\alpha=\frac{35-4\beta_0}{27}=\alpha_{\rm crit}$ have been adopted. Anisotropy profiles (see Appendix A for details) are described by the expression $\beta(r)=\beta_0+\beta_1\, (\gamma-\gamma_0)$, and are illustrated in the inset. Purple line refers to the reference profile in isotropic conditions with $\beta_0=\beta_1=0$, red line to $\beta_0=0$ and $\beta_1=0.2$, blue line to  $\beta_0=-0.1$ and $\beta_1=0.2$, and orange line to $\beta_0=+0.1$ and $\beta_1=0.2$.}\label{fig|NMC_density_anisotropy}
\end{figure}

For non-minimally coupled halos, an inner core with $\gamma_0\approx 0$ is enforced anyway by the second term in Eq.~(\ref{eq|fundeqpowlawanis}), so that the variations with respect to the isotropic case are minor and limited to the outermost regions. In Fig.~\ref{fig|NMC_density_anisotropy} we show how the non-minimally coupled density profile with $\eta=-0.05$, $\Gamma=\frac{5}{3}$ and $\alpha=\frac{35-4\beta_0}{27}=\alpha_{\rm crit}$ is affected by anisotropies. For realistic values $-0.1\lesssim \beta_0\lesssim 0.1$ and $\beta_0\approx 0.2$, the profile is marginally affected in the inner region and at intermediate radii, while it tends to extend toward slightly larger radii (i.e., the cutoff moves outward) due to the progressive prevalence of radial anisotropy in the halo outskirts; such an effect, though minor, is more pronounced for larger (more positive) $\beta_0$.

\end{document}